\documentclass[useAMS,usenatbib]{mn2e}

\usepackage{graphicx}

\usepackage{amsmath}
\newcommand{\lum}{erg s$^{-1}$}
\newcommand{\flux}{erg cm$^{-2}$ s$^{-1}$}
\usepackage{natbib}
\title[Hard X-ray emission of Sco X-1]{Hard X-ray emission of Sco X-1}
\author[Revnivtsev et al.]{Mikhail G. Revnivtsev$^{1}$\thanks{E-mail:
revnivtsev@hea.iki.rssi.ru},  Sergey S. Tsygankov$^{2,3,1}$,  Eugene M. Churazov$^{1,4}$, \newauthor Roman A. Krivonos$^{1,5}$\\
$^{1}$ Space Research Institute, Russian Academy of Sciences, Profsoyuznaya 84/32, 117997 Moscow, Russia\\
$^{2}$ Finnish Centre for Astronomy with ESO (FINCA), University of Turku,  V\"ais\"al\"antie 20,
FI-21500 Piikki\"o, Finland \\
$^{3}$ Tuorla observatory, Department of Physics and Astronomy, University of Turku, 
  V\"ais\"al\"antie 20, 21500 Piikki\"o, Finland \\
$^{4}$ Max-Planck-Institute f\"ur Astrophysik, Karl-Schwarzschild-Str. 1, D-85740 Garching bei M\"unchen, Germany\\
$^{5}$ Space Sciences Laboratory, University of California, Berkeley, CA 94720, USA\\
}
\begin{document}

\date{}

\pagerange{\pageref{firstpage}--\pageref{lastpage}} \pubyear{2013}

\maketitle

\label{firstpage}

\begin{abstract}
We study hard X-ray emission of the brightest accreting neutron star Sco X-1 with INTEGRAL observatory. Up to now INTEGRAL have collected $\sim4$ Msec of deadtime corrected exposure on this source. We show that hard X-ray tail in time average spectrum of Sco X-1 has a power law shape without cutoff up to energies $\sim200-300$ keV. An absence of the high energy cutoff does not agree with the predictions of a model, in which the tail is formed as a result of Comptonization of soft seed photons on bulk motion of matter near the compact object. The amplitude of the tail varies with time with factor more than ten with the faintest tail at the top of the so-called flaring branch of its color-color diagram. We show that the minimal amplitude of the power law tail is recorded when the component, corresponding to the innermost part of optically thick accretion disk, disappears from the emission spectrum. Therefore we show that the presence of the hard X-ray tail may be related with the existence of the inner part of the optically thick disk. We estimate cooling time for these energetic electrons and show that they can not be thermal. We propose that the hard X-ray tail emission originates as a Compton upscattering of soft seed photons on electrons, which might have initial non-thermal distribution.
\end{abstract}


\section{Introduction}
Sco X-1 is the brightest X-ray source on the sky. Its emission in X-ray energy band is powered by energy released in accretion of matter onto neutron star in a semi-detached binary system.
Sco X-1 receives a lot of attention because while studying such a bright source one can try to test theoretical models of the accretion flow with maximally possible precision.

The vast majority of X-ray emission of Sco X-1 (as well as of other neutron star binaries with high mass accretion rates) is formed in two parts of the accretion flow -- in optically thick accretion disk \cite[see model of][]{ss73} and optically thick layer between the accretion disk and the neutron star surface \cite[e.g.][]{mitsuda84}. Both these regions have temperatures about few keV and emit thermal spectra in 1-20 keV energy band.

In addition to these components, already early observations of Sco X-1 showed that sometimes an additional power-law like component is present at energies above 20-40 keV \citep{peterson66,agrawal71,haymes72,duldig83}, but sometimes - not \citep{lewin67,rothschild80,coe80,maisack94}. More recent studies of neutron star binaries at high mass accretion rates confirmed that amplitude of a hard X-ray power law component varies with time and likely correlates with detailed characteristics of spectral state of the source \citep{disalvo01,damico01,disalvo02,disalvo06,paizis06,dai07,sturner08,maiolino13}.

The origin of this tail remains unclear. Among different models of formation of this tail one can mention synchrotron emission of energetic electrons \cite[e.g.][]{riegler70}, Comptonizaion of seed photons on electrons of hot thermal corona, on non-thermal electrons \cite[e.g.][]{disalvo06}, or on bulk motion of matter near the compact object \cite[e.g.][]{paizis06,farinelli09}. In order to distinguish between these models one need to obtain high quality spectral energy distribution in hard X-rays because of possible high energy cutoff, predicted by some models, and to relate the behaviour of the hard X-ray tail to other components of X-ray emission of the source. In the present paper we will try to do this with the help of data of INTEGRAL orbital observatory.

At present the best instrument to study the hard X-ray/soft gamma-ray emission of Sco X-1 is INTEGRAL observatory \citep{winkler03}. This observatory has the best ever available sensitivity to hard X-ray emission, especially at energies above 150-200 keV.
During more than 10 years of operation this observatory performed more than 2000 separate observing sessions with combined exposure time (before dead time correction) about 5.7 Msec (including 4 Msec of observations, granted to authors of this paper). Sensitivity of the combined dataset reaches the level $\sim1$ mCrab at energies $\sim150$ keV and $\sim5$ mCrab at energies 200-300 keV.

\section{Data analysis and the sample}

We used all publicly available data of INTEGRAL observatory \citep{winkler03} on Sco X-1, including those, granted to authors of this paper (4 Msec of proposal ID 0720011). In total the dataset contains 2032 individual pointings. 

Two main instruments of INTEGRAL observatory: IBIS \citep{ubertini03,lebrun03} and SPI \citep{vedrenne03} cover a broad energy range from $\sim20$ keV to $\sim10$ MeV. We used data of instruments IBIS/ISGRI and SPI, which provide the best sensitivities at energies 20--500 keV. X-ray monitor JEM-X of INTEGRAL observatory has field of view $\sim4.8^\circ$ much smaller than those of IBIS and SPI ($\sim9^\circ$ and $\sim16^\circ$ correspondingly), and therefore have collected much less exposure on Sco X-1 (observations of INTEGRAL is composed of a set of individual pointings -- dithering pattern -- which span 10 degrees in both directions). Therefore in order to extend our energy coverage down to 3 keV used used simultaneous observations with RXTE observatory.

\subsection{IBIS/ISGRI data analysis}
\label{isgri}
Total exposure of IBIS/ISGRI observations is $\sim5.7$ Msec before deadtime correction and $\sim3.8$ Msec after deadtime correction.

INTEGRAL/IBIS/ISGRI data were processed with the method described in \cite{revnivtsev04,krivonos07,krivonos10}. The spectral characteristics of the ISGRI detector changes with time, which leads to variations of the spectral gain (conversion of instrumental channels to photon energies) and energy resolution of the instrument \cite[see e.g.][]{caballero13}. In order to put data of all ISGRI observations into proper energy scale we have performed gain corrections for all individual observations with the help of fluorescence line of tungsten at $\sim59$ keV characteristic of the ISGRI background \citep{terrier03,tsygankov06,caballero13}. Therefore, at energies around 60 keV we can estimate an uncertainty of channel-to-energy conversion to be less then $\sim0.5$ keV, while at lower energies 20-30 keV this uncertainty can increase up to $1-2$ keV, or $\sim 1$\%. Thus, in all our subsequent analysis we assume that the accuracy of our values, related with the energy scale is not better than $\sim 1$\%.

For all observations we have produced a set of images in a number of energy intervals. 
The raw count rate of sources determined from these images was corrected for two main effects: offset-angle dependent absorption of X-rays in IBIS mask support structure and long term variations of the instrument characteristics. These effects were measured with the help of numerous observations of Crab nebula, which we consider to be a stable source. Correction for absorption in the IBIS mask support structure was done with the help of simple axi-symmetric analytic function, fitted to the Crab nebula observations, performed at different offsets. Time dependent count rate correction was done using Crab nebula count rates, averaged over 400 days time bins. 

The resulting light curves of the Crab nebula, corrected for these two main effects, nevertheless still have residual variances exceeding those caused by a Poisson statistics. In energy band 17-26 keV  the residual rms variations have amplitude 5.7\%, in energy band 26-38 keV 4.2\%, in energy  band 38-57 keV 4.2\%, in energy band 57-86 keV 4.5\%. At higher energies the rms variations of the corrected count rates, measured in individual pointings ($\sim1-2$ ksec exposure time) are already compatible or less than those created by Poisson statistics. In our study we treat the remaining residuals as being artificial due to various unaccounted instrumental effects, while existence of small (at the level of a few percent) intrinsic variations of the Crab nebula hard X-ray flux is still possible \cite[see e.g. study of Crab nebula flux variations in ][]{wilson11}. The limited accuracy of the inferred Crab nebula flux stability means that our sensitivity to Sco X-1 spectral variations is limited by these systematic uncertainties, which we should keep in mind in the subsequent analysis. All model fits to INTEGRAL data points take into account these additional systematic uncertainties at the level of 4-5\%.

\subsection{SPI data analysis}

In addition to the ISGRI data we utilized data obtained with the SPI spectrometer onboard INTEGRAL observatory. It operates in energy range from 20 keV to 8 MeV and complements the IBIS telescope in studies of hard emission from compact objects. Coded aperture mask gives to the spectrometer imaging capabilities with an angular resolution of 2.5$^\circ$(FWHM). 

Primary data screening and reduction was done following procedures described in \cite{churazov05,churazov11}. 

The model used to derive the spectrum consists of a single point source (at the position of Sco X-1) and a constant (two free parameters for each energy channel). For individual observations the response to a point source is calculated using Instrument Response Functions (IRFs). This model is then applied to the background subtracted SPI data and the best fitting fluxes from the source in narrow energy channels form the Sco X-1 spectrum.

To reduce an undesirable contribution from the Galactic diffuse emission and other point sources located in the Galactic Centre region we selected only observations, in which the center of the SPI field of view was within 10$^{\circ}$ circle around Sco X-1. The total net exposure time of used SPI observations is about 4.3 Msec. 

Inclusion to the sky model additional point sources in the SPI field of view (1E 1740.7-294, GX 1+4, 4U 1700-377, MAXI J1659-152) and Galactic diffuse emission does not affect significantly the resultant spectrum of Sco X-1.

\subsection{RXTE/PCA data analysis}

In order to extend the energy spectrum of the source down to 3 keV we have studied spectra of Sco X-1, obtained by spectrometer PCA of RXTE observatory \cite{bradt93} during time periods of INTEGRAL observations (similar to those in works of \citealt{disalvo06,sturner08}).  All data were analysed with standard tasks of HEASOFT v6.15. 

\section{Two patterns of flux variations}

\begin{figure}
\includegraphics[width=\columnwidth,bb=16 183 570 720,clip]{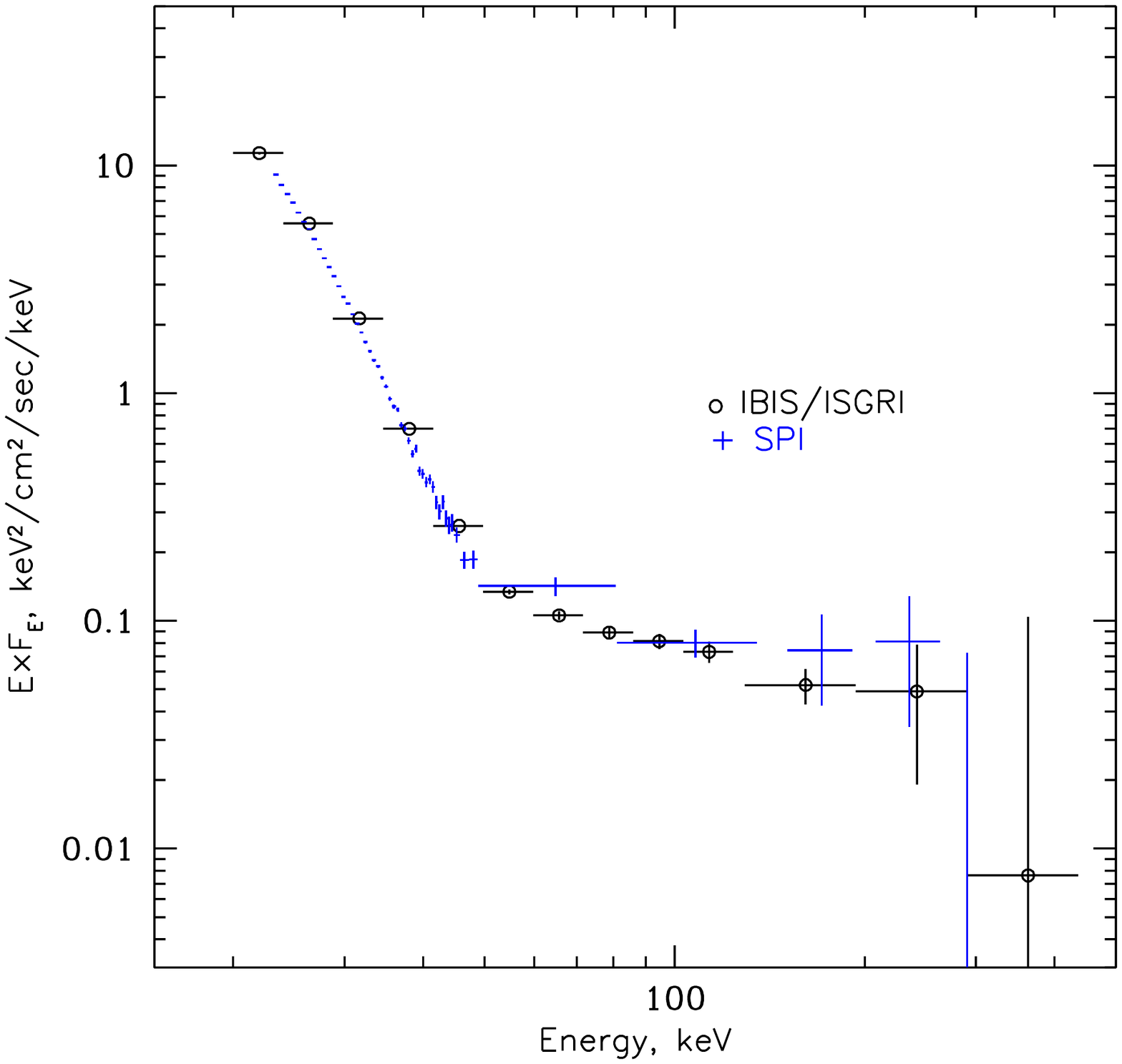}
\caption{Spectrum of Sco X-1 in hard X-rays measured by INTEGRAL instruments, averaged over all observations. Open circles denote spectrum, measured by IBIS/ISGRI instrument, crosses - with SPI.}
\label{sp_aver}
\end{figure}

The spectrum of Sco X-1, averaged over all analysed observations is presented in Fig.\ref{sp_aver}. Spectra, measured by two INTEGRAL instruments are perfectly consistent with each other.

\begin{figure}
\includegraphics[width=\columnwidth,bb=16 183 570 620,clip]{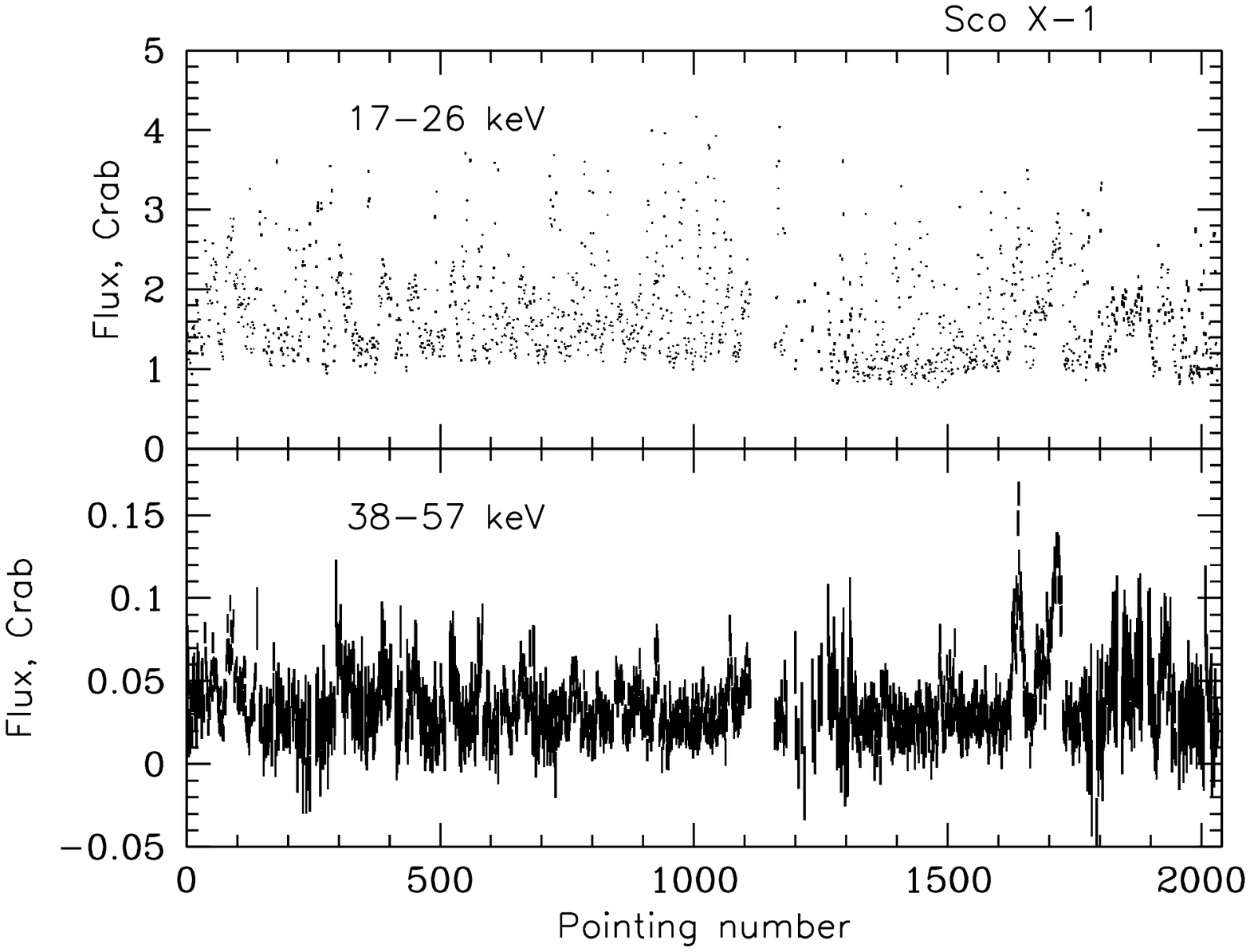}
\caption{Light curve of Sco X-1 as measured by IBIS/ISGRI in units of Crab nebula count rate in the same energy ranges. X-axis here shows the running number of INTEGRAL pointings. We present the light curve in such a way because of big gaps between different sections of observations. }
\label{lcurve_total}
\end{figure}

Long term light curve of Sco X-1 is shown in Fig.\ref{lcurve_total} (the source fluxes here and below are shown in units of Crab nebula count rate in appropriate energy ranges). It is seen that the hard X-ray flux of Sco X-1 is strongly variable, at the level far exceeding instrumental uncertainties (see description of systematic in Sect. \ref{isgri}).

\begin{figure}
\vbox{
\includegraphics[width=\columnwidth,bb=6 183 570 620,clip]{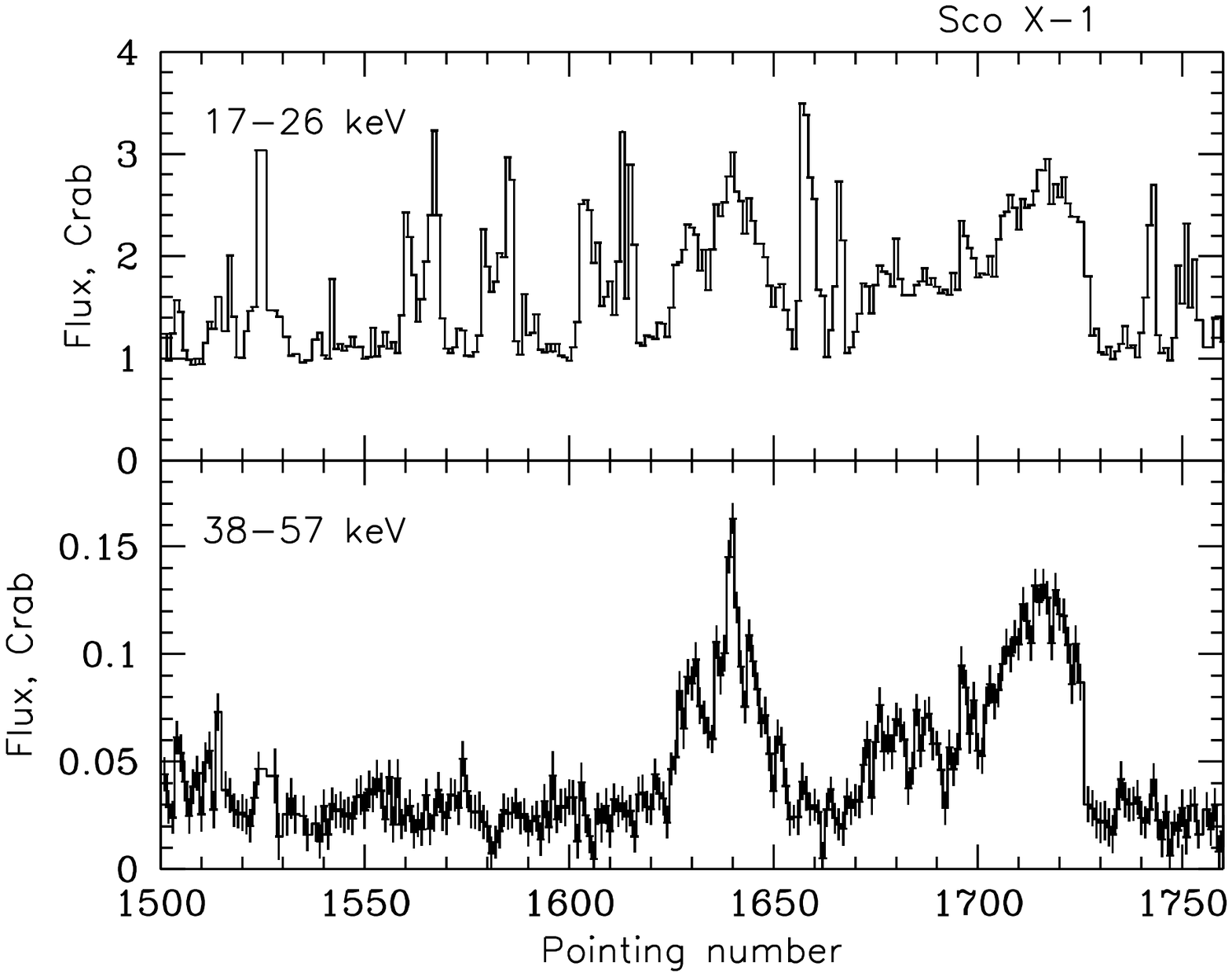}
\includegraphics[width=\columnwidth,bb=6 183 570 720,clip]{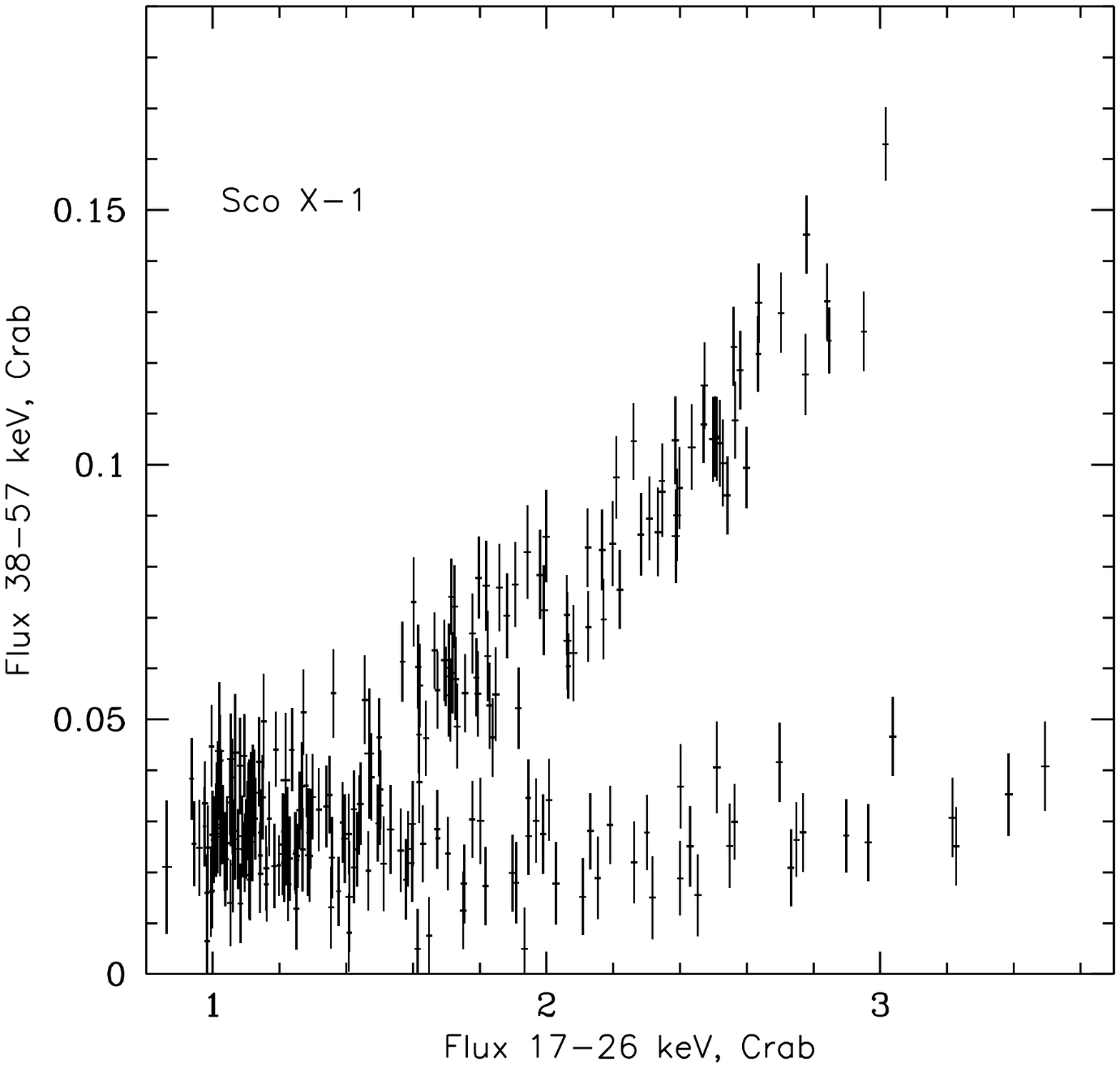}
}
\caption{Upper panel: light curve of Sco X-1 in energy bands 17-26 keV and 38-57 keV during some part of observations, selected around hard X-ray flares. Lower panel: flux-flux diagram of these lightcurves.}
\label{lcurve_flare}
\end{figure}

Time variability of Sco X-1 fluxes has two branches. We can see that strong variability of the source flux in 17-26 keV energy range can be either accompanied by similar variability at harder X-rays (38-57 keV) or not. The spectacular example of such bi-modality can be seen at data points closer to the end of our dataset. The close-up to this set of points is shown in Fig.\ref{lcurve_flare}.

\begin{figure}
\includegraphics[width=\columnwidth,bb=6 183 570 720,clip]{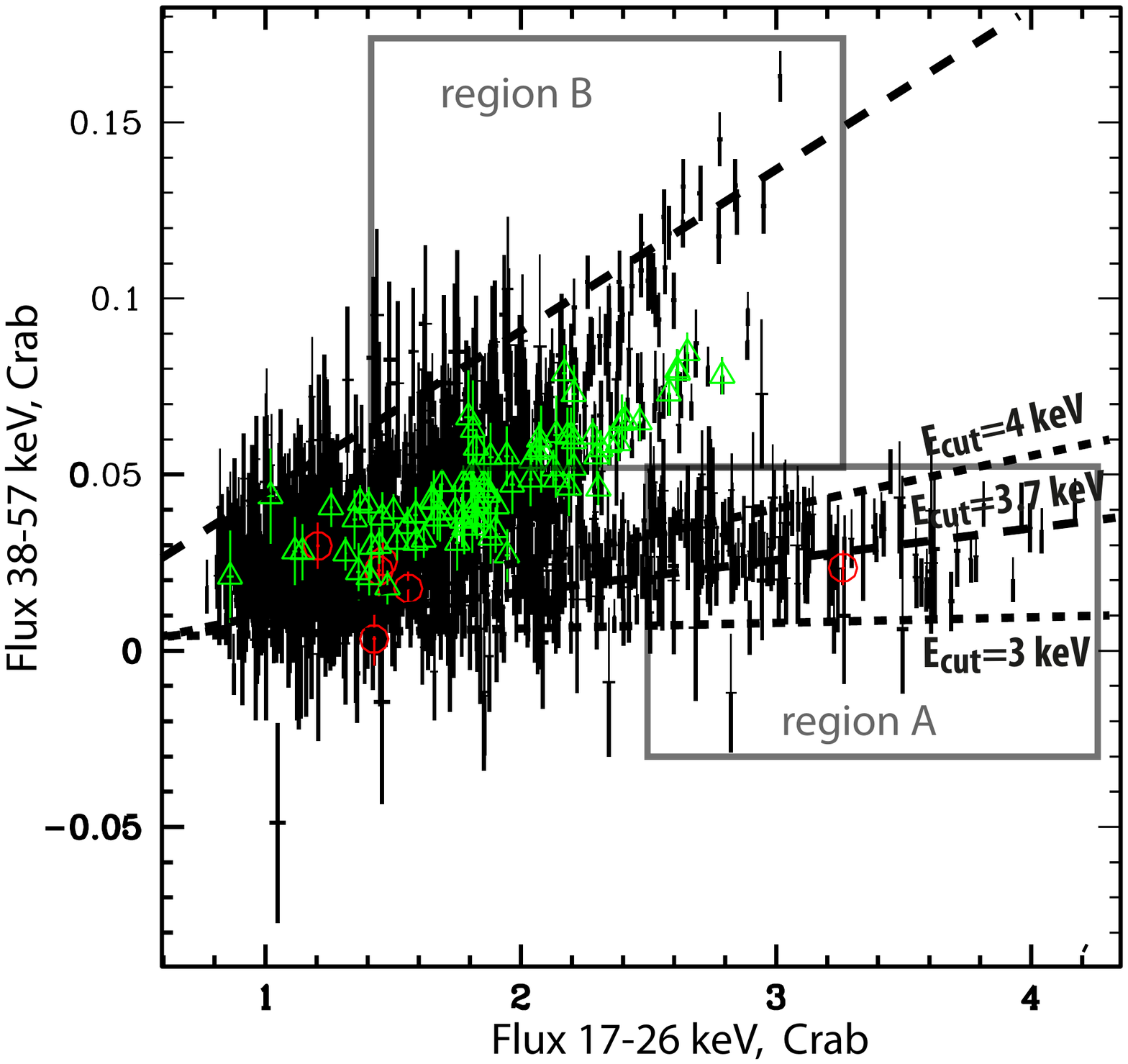}
\caption{Flux-flux diagram of light curves of Sco X-1 in energy bands 17-26 keV and 38-57 keV, presented for all observational dataset. Open triangles and circles show fluxes of Sco X-1, measured during observations, performed simultaneously with RXTE. Open triangles denote observations in horizontal and normal branches, open circles - on flaring branch. Dashed line shows correlation of two fluxes, which corresponds to spectral shape $dN/dE\propto \exp(-E/E_{\rm cut})$ with $E_{\rm cut}=3.7$ keV, appropriate for emission of the boundary/spreading layer on NS surface.  Upper and lower dotted lines show correlation $E_{\rm cut}=4$ keV and $E_{\rm cut}=3$ keV respectively.}
\label{ffd}
\end{figure}

\begin{figure}
\includegraphics[width=\columnwidth,bb=20 175 571 700,clip]{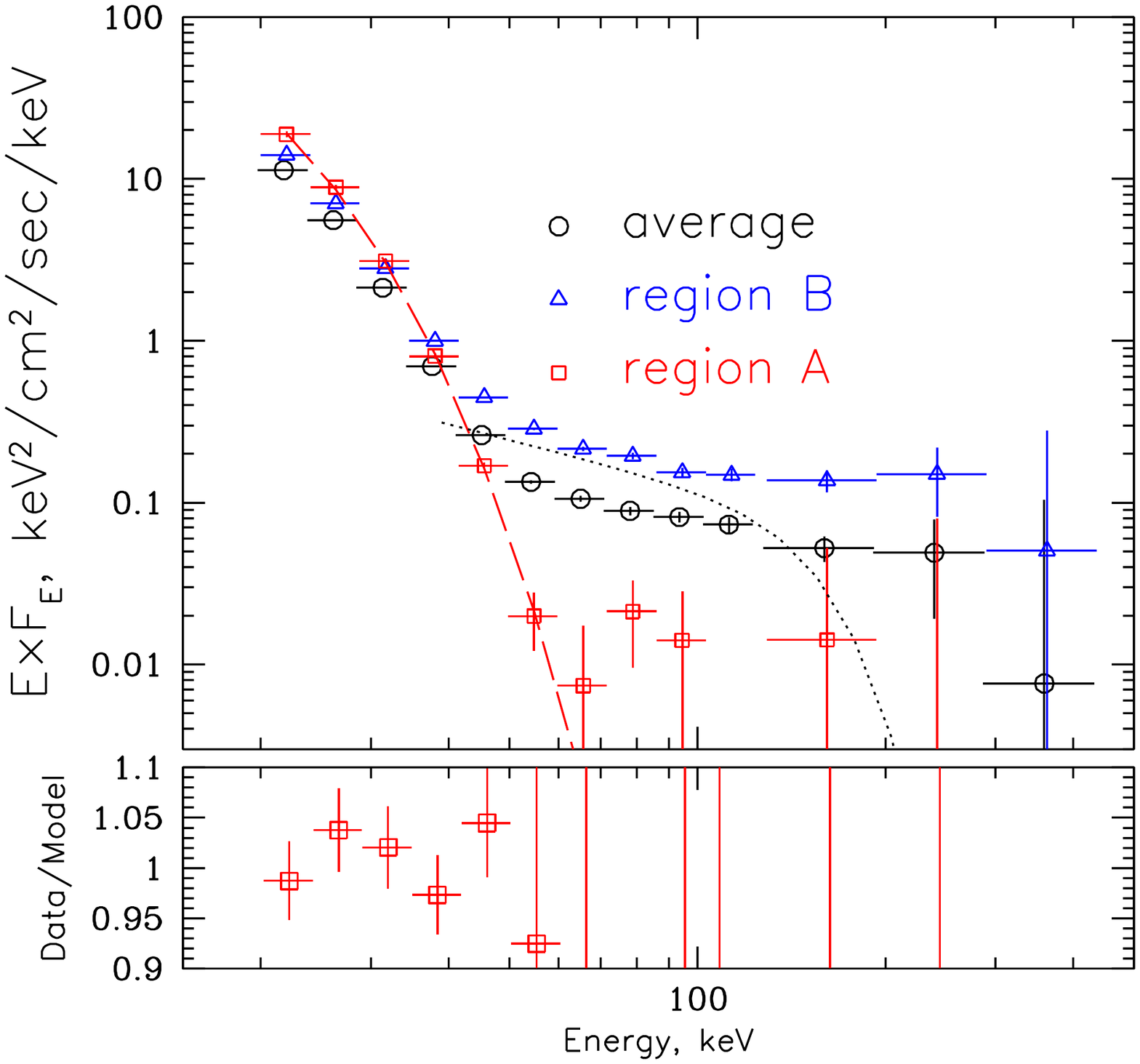}
\caption{Spectra of Sco X-1, collected during different time periods. Open circles denote time averaged source spectrum, open triangles - spectrum collected within sector A of Fig.\ref{ffd}, open squares - spectrum, collected within sector B of Fig.\ref{ffd}. Dashed curve shows simple analytic model $dN/dE\propto \exp(-E/3.7\textrm{keV})$, representing emission of the boundary/spreading layer on NS surface. Dotted curve - simple analytical fit (following receipt of Zdziarski et al. 2001) to Monte Carlo simulation of emission Comptonized by bulk motion of matter around black holes, Laurent \& Titarchuk (1999). Bottom panel shows data to model ratio of maximally tail-free spectrum of Sco X-1 (red open squares). It is seen that up to statistically significant points at 60 keV the data to model ratio is within our adopted systematic uncertainties of $\sim4-5$\%.}
\label{sp_tails}
\end{figure}

Spectra, illustrating the two branches are shown in Fig.\ref{sp_tails}.

 The spectrum collected in observations, occupying the lower right corner of the flux-flux diagram (region A in Fig.\ref{ffd}), has virtually no power law tail and can be adequately fitted with simple thermal component. The best fit analytical approximation can be done by an exponential cutoff $dN/dE\propto E^{0}\exp(-E/E_{\rm cut})$ with $E_{\rm cut}=3.70\pm0.03$ keV or blackbody emission with $kT=2.8\pm0.05$ keV. 
 
The upper right part of the diagram is formed by spectra, with significant power law tail.
Spectrum collected within this region (region B) is shown by open triangles in Fig.\ref{sp_tails}. The hard X-ray part of this spectrum (above 60 keV) can be described by a power law with photon index $\Gamma=2.6\pm0.3$ without high energy cutoff. The exact value of the lower limit on the cutoff energy depends on assumed functional shape of the model. For simple exponential cutoff the lower limit (2$\sigma$) is $E_{\rm cut}>330$ keV.  The 50-250 keV flux in this tail is $\sim 4-5 \times10^{-10}$ \flux .

\section{Relation of the tail with other spectral components}

It was previously shown \citep{disalvo01,disalvo06,dai07} that the presence of the tail in spectra of bright NS binaries is related with position of sources on color-color diagram (CCD). The maximally luminous tail is observed on so-called horizontal and normal branches of CCD \cite[see e.g][for description of CCD phenomenology]{hasinger89} and the weakest power law tail - on the flaring branch.

While physical origin of different branches on CCD of bright neutron star binaries is not yet understood in details, several important clues are already obtained. 

Spectral decomposition of emission of luminous neutron star binaries is a complicated issue due to similarity of spectral shapes of two main components: optically thick accretion disk and a boundary/spreading layer \cite[see e.g. different decompositions in][]{disalvo02,dai07}. In our work we adhere approach of model independent decomposition, based on properties of variability of luminous neutron stars, outlined in works of \cite{gilfanov03,revnivtsev06}.

In these works it was shown that energy spectra of bright neutron star binaries (at least in energy band 1-20 keV) may be decomposed into contribution 
of optically thick accretion disk with temperatures $kT\sim1-1.8$ keV and the neutron star surface with color temperature $kT\sim2.5-2.7$ keV. The latter has virtually constant spectral shape over the whole range of luminosities and spectral hardnesses unless the source transit into optically thin regime of emission \citep{mitsuda84,gilfanov03,revnivtsev06}. 

Stability of the hotter thermal component spectral shape can be understood as Eddington limited emission of optically thick radiation pressure dominated spreading/boundary layer, where matter settles from rapid rotation in accretion disk to slower rotation of the neutron star \citep{inogamov99,suleimanov06}

Variations of hardness (colors) of observed X-ray emission of bright NS binaries in energy band 2-20 keV occur due to variations of the accretion disk temperature and its fractional contribution to the total X-ray emission.

It was shown that at the top part of the flaring branch the X-ray emission of bright NS binaries consists of almost solely the neutron star surface/boundary/spreading layer \citep{revnivtsev13}. Contribution of the innermost parts of optically thick accretion disk drops significantly to almost undetectable levels while the total source brightness might increase in this state. Origin of such a decrease of the accretion disk component contribution is not clear yet and requires further study. 

 We should emphasize again  here that this conclusion is based on spectral decomposition, which uses information about the source flux variability. The decompositions based on spectral modelling only may provide different results (see e.g. \citealt{dai07}).

\begin{figure}
\includegraphics[width=\columnwidth,bb=20 183 570 720,clip]{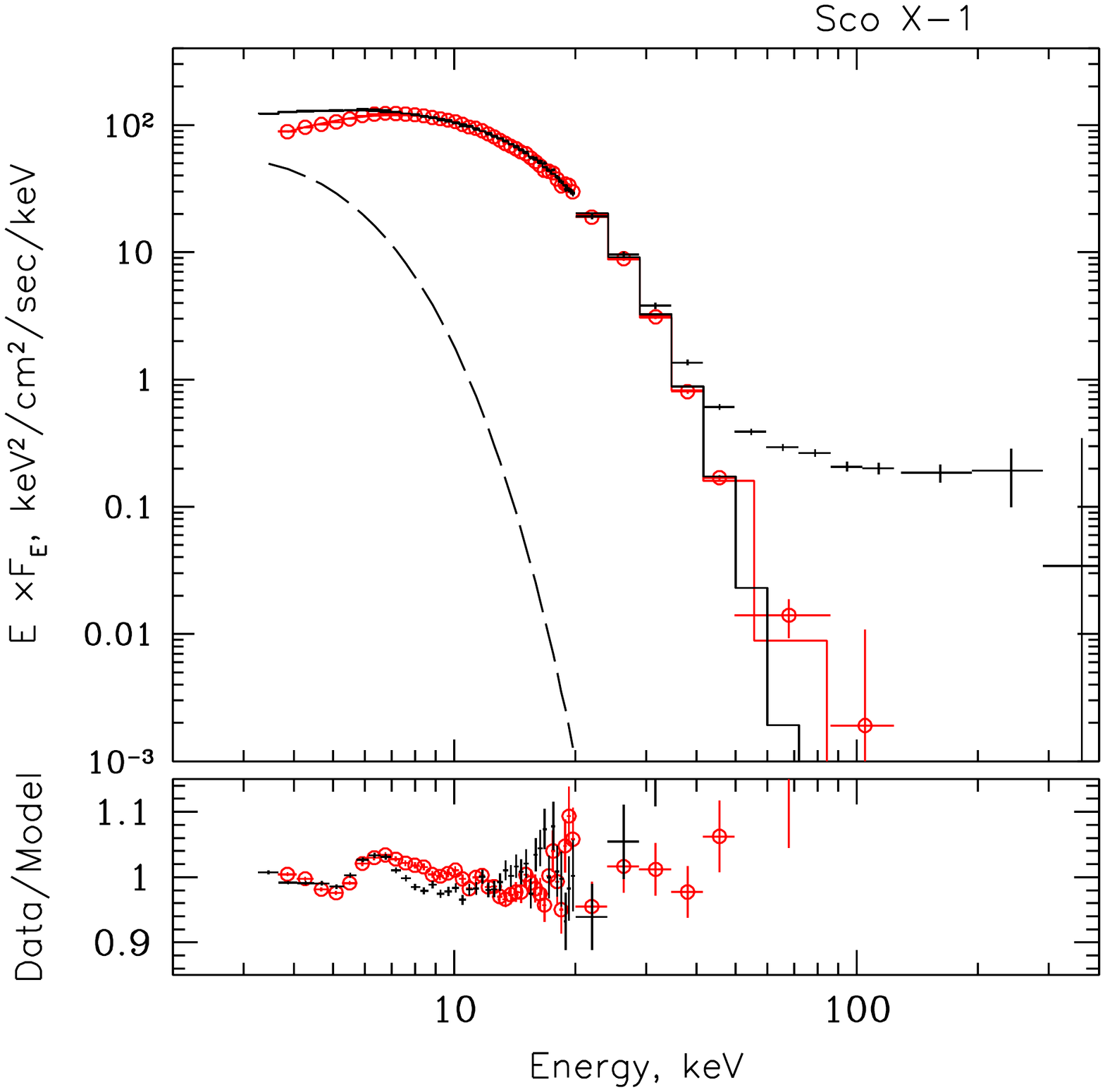}
\caption{X-ray spectra of Sco X-1 at two spectral states. Red crosses show spectrum collected on flaring branch of its CCD evolution, INTEGRAL/IBIS data collected over region A in Fig.4, at softer X-rays we use RXTE/PCA data at the top of the flaring branch (128 sec spectrum). Black crosses show spectrum at horizontal branch of Sco X-1, INTEGRAL/IBIS data collected over region B of Fig.4, at softer X-ray we use RXTE/PCA data (128 sec) within time interval shown by topmost triangle on Fig.4. Spectra are scaled vertically to match at energies 15-25 keV.
Red solid curve shows analytic model of boundary/spreading layer on NS surface (see text), long-dashed curve shows multicolor black body disk component. It is seen that disappearance of the power law tail in spectrum is accompanied by disappearance of multicolor disk component. Bottom panel shows the ratio of the data points to the simple model, consisting of contribution of the accretion disk and the boundary/spreading layer. The iron fluorescent emission line at $\sim6.4$ keV, not included in the model, is clearly seen in the residuals.}
\label{2sp_fb}
\end{figure}

As an illustration of such behavior of Sco X-1 we present Fig.\ref{2sp_fb}. As the accretion disk contribution in energy spectrum is visible only at energies below 10 keV we can not limit ourself with data of IBIS and SPI instruments of INTEGRAL observatory (operating at energies $>20$ keV). In order to reach energies about 3 keV and to maximize the statistics we used data of simultaneous observations of Sco X-1 with IBIS/INTEGRAL and PCA/RXTE. Requirement of simultaneity of INTEGRAL and RXTE observations leaves only a small fraction of all available datasets.

 At  Fig.\ref{2sp_fb} we present broad band spectra collected at the top and the bottom parts of the "flaring branch" (128 sec exposure time of RXTE/PCA at energies 3-20 keV). Both spectra were fitted as a sum of boundary/spreading layer component (we adopted simple analytic form $dN/dE\propto E^{0} \exp(-E/3.7 \textrm{keV})$ following \citealt{gilfanov03,revnivtsev06,revnivtsev13}) and multicolor disk blackbody component \citep{ss73,mitsuda84}. We see that while at the bottom of the flaring branch the multicolor disk contribution is comparable (or even larger) than that of the boundary/spreading layer, on top of the flaring branch this contribution is absent.

Therefore, the difference of the top part of the flaring branch from any other branches of Sco X-1 is the absence of the spectral component attributed to the innermost optically thick accretion disk. Note, that similar (in terms of spectral component analysis) disruption of the inner disk was previously observed in black hole candidates GRS 1915+105 \citep{belloni97}.

Study of hard X-ray emission of Sco X-1 shows that the power law hard X-ray tail also disappears at this stage. On Fig.\ref{ffd} flaring branch observations are denoted by open circle.

 From this plot we can make two important conclusions:

\begin{itemize}
\item Contribution of hard X-ray power law tail is small at the top part of the flaring branch (similar conclusion was previously obtained by \citealt{disalvo06,sturner08})
\item Variations of fluxes in energy bands 17-26 keV and 38-57 keV on the top of the flaring branch (i.e. without power law tail) occur with constant spectral shape (the straight line best fit to the data points within region A on flux-flux diagram of Fig.\ref{ffd} has a reduced $\chi^2/d.o.f\sim1.18$, while within region B it has a value $\chi^2/d.o.f\sim3.06$ ). This conclusion strongly supports our previous findings \citep{gilfanov03,revnivtsev06,revnivtsev13} that the hotter thermal component, which we ascribe to emission of the boundary/spreading layer on NS surface, preserve its shape over all flux variation of bright NS binaries. These hard X-ray observations are important because at these energies $E>10kT$ our sensitivity to variations of the color temperature of the thermal component is exponentially high (at energies $E>>kT$ even small variations of the color temperature of the emission lead to large variations of the spectral hardness). 
\end{itemize}

Making use our variability-based spectral decomposition \cite{gilfanov03,revnivtsev06} basing on abovementioned arguments we conclude that the hard X-ray power law tail in spectrum of Sco X-1 may be related with existence of the innermost part of the optically thick accretion disk and not with hotter thermal component (boundary/spreading layer emission). This might reflect difference in energy release mechanisms in MRI-dominated accretion disk and in the slowly decelerating spreading layer on neutron star surface \citep{inogamov99,inogamov10}. It is expected that in the former case powerful corona should form \cite[see e.g.][]{miller00,schnittman13} , while it might not form in the latter case.

\section{Physical origin of the tail}

The vast majority of X-ray emission of Sco X-1 is generated in optically thick accretion disk and boundary/spreading layer. Hard X-ray tail, which does not have an exponential cutoff up to at least 200-300 keV, should be generated within some optically thin region with energetic electrons and the tail should not be directly associated with the emission of NS surface (the tail is not seen during time periods with exclusive contribution of the NS surface/boundary/spreading layer).
In this case it is reasonable to seek for analogies with the emission mechanisms of accreting stellar mass black holes.

Studies of emission of accreting black holes showed that optically thin regions with energetic electrons exist in two flavours \cite[see e.g. review in][]{done07}. First --  as optically thin part of the accretion flow inside some distance from the compact object, where the optically thick disk is truncated, the hot electrons in this region are thermalized and the emitting spectrum can  be described by Comptonization on thermal electrons \citep{sle76,sunyaev79,poutanen97}. Second -- as region, where hard X-ray/soft gamma-ray tail in energy spectra of bright accreting black hole systems does not show evidences of thermal cutoff. The widely accepted interpretation of this emission is a Compton upscattering of some seed photons on hybrid/non-thermal electrons \cite[see e.g.][]{gierlinski99,coppi99,zdziarski01}. 

Among other mechanisms which were proposed for generation of hard X-ray/soft-gamma-ray tail in emission spectra of accretion disks one can mention Comptonization on bulk motion of electrons, moving with relativistic speed close to the black hole horizon \citep{mastichiadis92,titarchuk97,laurent99}. In some studies similar mechanism was also considered for accreting neutron stars \citep[see e.g.][]{paizis06,farinelli09}. 

The capability of this mechanism to produce the observed hard X-ray tails in spectra of accreting black holes was discussed in details in \cite{zdziarski01}. It was shown that the main contradiction of bulk motion Comptonization models with the data is inevitable rollover of their predicted energy spectra in hard X-rays. The cutoff energy in the spectrum depends on the bulk radial velocity of the matter close to the compact object and can not be higher than $E_{\rm cut}\sim m_e c^2/\textrm{few}\sim100-200$ keV \cite[see e.g][]{laurent99,zdziarski01,farinelli12}. The most sensitive observations of some black hole binaries show a power law tail without noticeable cutoff up to energies 500-600 keV. 

In the case of Sco X-1 we have statistically significant spectral data points up to $\sim200$ keV. In order to visualize the spectral shape of the bulk Comptonization model (for the extreme case of accreting black hole from work of \citealt{laurent99}) we present dotted curve on Fig.\ref{sp_tails}. This curve represents analytic approximation (from \citealt{zdziarski01}) of Monte Carlo simulated bulk motion Comptonization spectrum. It is clear that even in this case of assumed accretion onto black hole, in which the bulk motion velocities of matter are significantly higher than those in the case of accreting neutron stars, the spectrum has a cutoff at $\sim150$ keV and this cutoff strongly contradicts the observed spectral data points.

For pure free fall onto neutron star with radius $R_{\rm NS}=3R_g$ the maximum radial velocity of the matter corresponds to $\beta=v_r/c<0.6$, which is already smaller than that near black holes. In the case of accreting neutron stars with emerging luminosities close to the Eddington limit  (e.g. Sco X-1) the bulk radial velocity of the matter should be even smaller than that.  This should result in even smaller cutoff energy in the spectrum of the tail \cite[see e.g. numerical models in][]{farinelli12}. 

{\sl Therefore we can conclude that obtained hard X-ray spectrum of Sco X-1 does not agree with the predictions of a model, in which the hard X-ray tail is formed as a result of Comptonization of some seed photons on bulk motion of the matter around accreting compact object.}

Thermal Comptonization of seed photons in hot (with temperatures above 100 keV, which might create an observed power law tail up to energies $>200-300$ keV) optically thin inner part of the accretion flow does not seem to be probable because huge soft photon flux from neutron star surface ($L_{\rm soft}>10^{38}$ \lum ) with Compton temperature around few keV very effectively cools down (via Compton cooling) any possible hot ($\ga100$ keV) thermal flow.

Compton cooling time for energetic electron can be estimated as:

$$
t_{\rm Compt}\la \pi \gamma^{-1} l_{s}^{-1} {R\over{c}}
$$

where $l_s$ is the soft emission compactness:

$$
l_{s}={L_{\rm soft}\over{R}}{\sigma_T\over{m_e c^3}}>600
$$

Here we adopted that the size the emitting region $R$ does not exceed few NS radii (because we assume that it is related with the innermost parts of the accretion flow), i.e. $R\la 5 \times10^{6}$ cm. We see that the Compton cooling time for electrons is significantly less than the light crossing time of the emitting region. 

Thermalization time for electrons in this region depends on processes, which determine the energy exchange. Coulomb scattering of relativistic electrons of energies $\gamma m_e c^2$ with thermal bath of non relativistic electrons thermalize them on time-scale \cite[see e.g.][]{stepney83,svensson99}
$$
t_{\rm Coulomb}\sim {\gamma^2 \over{\tau_{\rm Th} \ln \Lambda}} \left({R\over{c}}\right)
$$
that is larger than $t_{\rm Compt}$ in our case of relativistic electrons in optically thin $\tau_{\rm Th}<1$ region. Here $\ln \Lambda$ is the Coulomb logarithm and $\tau_{\rm Th}$ -- Thomson optical depth of the region. Exchange of energies between electrons due to synchrotron self-absorption (so-called synchrotron boiler mechanism, \citealt{ghisellini88,svensson99,malzac09}) in a region with magnetic compactness $l_{\rm B}$:

$$
l_{\rm B}={\sigma_{\rm T}\over{m_e c^2}} R {B^2\over{8\pi}}
$$ 

gives thermalization time:

$$
t_{\rm synch}\sim \gamma^{-1}l_{\rm B}^{-1}  {R\over{c}} 
$$
, which in our case ($l_{s}\gg l_{\rm B}$) is larger than Compton cooling time scale.

\begin{figure}
\includegraphics[width=\columnwidth,bb=132 46 689 564,clip]{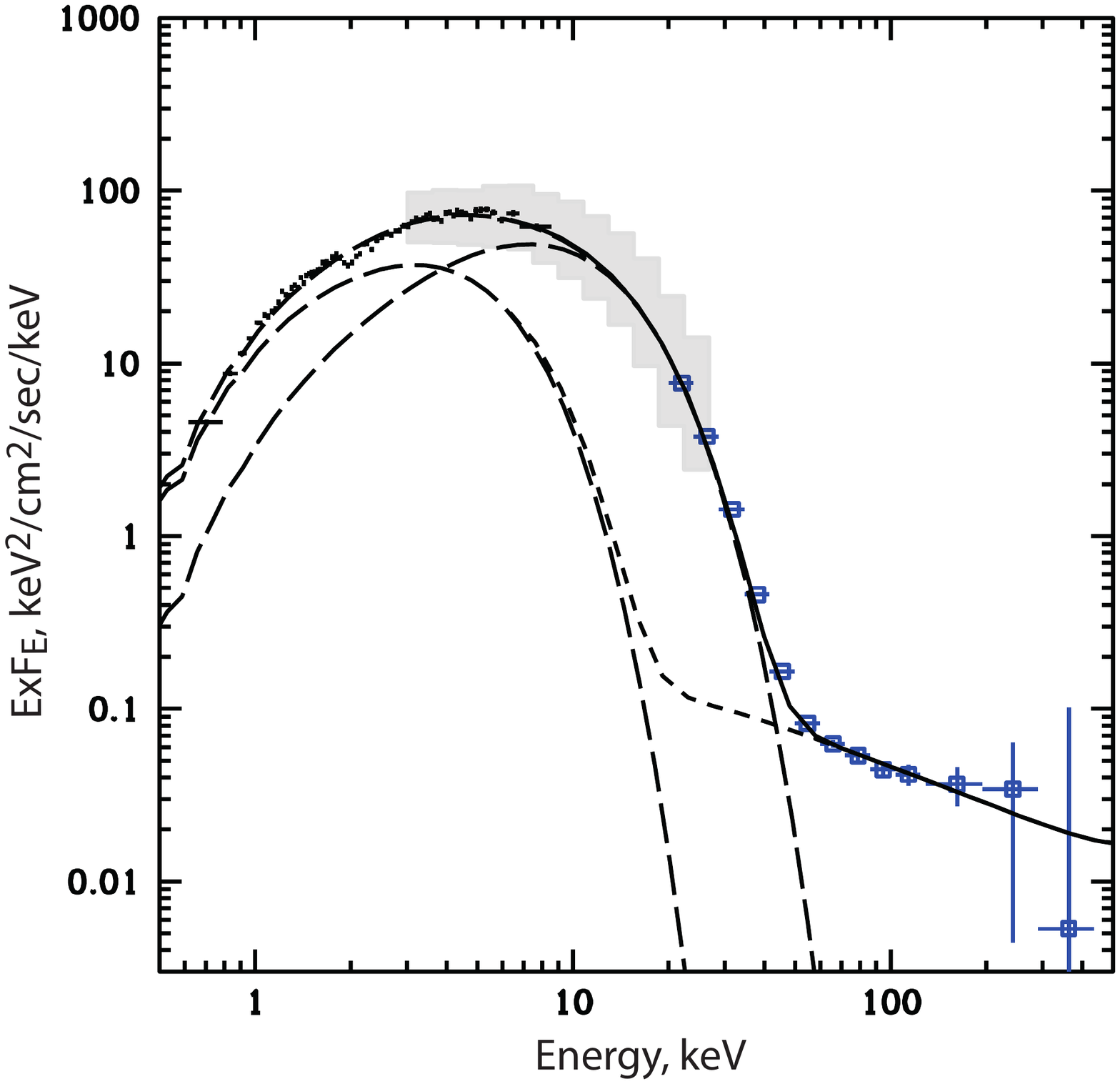}
\caption{Broadband energy spectrum of Sco X-1. At energies below 7 keV crosses show data points, measured with ASCA observatory (observation Aug.16, 1993), at energies 3-20 keV gray stripe denotes the rms-range of spectral densities, measured by RXTE/PCA over around 700 observations, collected during period 1996-2012. At energies above 20 keV open squares show time averaged spectrum of Sco X-1, measured with INTEGRAL/IBIS/ISGRI. Long dashed curves denote contribution of an accretion disk (left curve) and a boundary/spreading layer (right curve) components. Short-dashed curve shows prediction of a model, in which soft blackbody photons with $kT=1.1$ keV are Comptonized by hybrid/non-thermal plasma (model $eqpair$, see text for other parameters). Solid line is a sum of all components}
\label{broad}
\end{figure}

We can conclude that in the absence of some special collision-less plasma mechanisms of thermalization the optically thin hot plasma in the innermost accretion flow can not be uniformly thermal because the electrons do not have time to thermalize. The hard X-ray photons emerging in the tail (above 50-60 keV) should be result of Compton upscattering of soft photons on electrons having their original distribution. This original distribution can be non thermal \cite[see e.g. simulation of reconnection events,][]{zenitani01,yamada10}, depending on the physical mechanism of energy input into electron population.

We can assume that one of the most probable scenario of generation of hard X-ray tail in spectrum of Sco X-1 is a non-thermal Comptonization in some corona-like regions above the accretion disk \cite[see e.g. scenario of][]{galeev79}. Such scenario have direct analogy with accreting black holes in so-called soft state. According to widely accepted model during this state the accretion to a black hole occurs via optically thick disk with optically thin corona above it. Optically thick disk emits blackbody-like emission, while optically thin corona with energetic electrons upscatters these soft photons to higher energies \cite[see e.g.][]{coppi99,gierlinski99,done07}. This optically thick disk should be very similar to that around neutron star in Sco X-1, therefore we can anticipate that somewhat similar corona-like flow above this disk can also exist in this case.

An example of the spectrum, expected in such hybrid thermal/non-thermal plasma (model $eqpair$ of \citealt{coppi99} in XSPEC package, \citealt{arnaud96}) is presented in Fig.\ref{broad}.

For this energy spectrum we assumed that the hybrid/non-thermal plasma upscatters only photons from accretion disk and adopted the following parameters: $l_{\rm nth}=0.99$ fraction of energy deposited into non-thermal electrons with the power law slope $\Gamma=3$, soft photons luminosity compactness $l_s=600$, fraction of the energy, inserted into hard component with respect to the soft component $l_h/l_s=0.6$\%, temperature of the soft (black body) photons emission $kT_{bb}=1.1$ keV.

We should note here that energetic electrons in this optically thin corona can upscatter not only photons from optically thick accretion disk, but also photons from the neutron star surface (boundary/spreading layer). This should result in modest changes of the shape of the spectral model, shown in Fig.\ref{broad}, but more quantitative description of this modification will depend on  not known geometry of the Comptonizing region and is beyond the scope of the present paper.

\section{Summary}

We have studied all available data of INTEGRAL observatory on hard X-ray emission of the brightest accreting neutron star Sco X-1. In total it sum up to $\sim6$ Msec of astronomical time and $\sim4$ Msec of deadtime corrected exposure time.
Our results can be summarized as follows:

\begin{itemize}
\item Time average spectrum of Sco X-1 contains hard X-ray tail which has a power law shape without cutoff up to energies $\sim200-300$ keV.
\item This allows us to conclude that obtained hard X-ray spectrum of Sco X-1 does not agree with the predictions of a model, in which the hard X-ray tail is formed as a result of Comptonization of some seed photons on bulk motion of the matter around accreting compact object
\item The amplitude of the tail varies with time with factor more than ten.
\item We confirm previous findings that amplitude of the hard X-ray tail in energy spectrum of Sco X-1 correlates with its position on color-color diagram. The faintest tail is seen at the top of the so-called flaring branch.
\item  Our spectral decomposition of the energy spectrum of Sco X-1 shows an absence (or significant decrease of contribution) of optically thick accretion disk component at the top part of the flaring branch. Therefore we conclude that the presence of the hard X-ray tail may be related with the existence of the inner part of the optically thick disk.
\item Assuming that hard X-ray tail is generated above the innermost part of optically thick accretion disk we estimate cooling time for energetic electrons and show that they can not be uniformly thermal. We conclude that the hard X-ray tail emission originates as a Compton upscattering of soft seed photons on electrons, which have intrinsically non-thermal distribution.
\end{itemize}

\section*{Acknowledgements}

Research is based on observations with INTEGRAL, an ESA project with
instruments and science data centre funded by ESA member states (especially the
PI countries: Denmark, France, Germany, Italy, Switzerland, Spain),
Czech Republic and Poland, and with the participation of Russia and the
USA. Research has made use of data obtained
through the High Energy Astrophysics Science Archive Research Center
Online Service, provided by the NASA/Goddard Space Flight Center.
Authors thank Max Planck Institute fuer Astrophysik for computational support.
The work was supported by Russian Scientific Foundation (RNF), project 14-22-00271.

\end{document}